\documentclass{article}
\usepackage[a4paper,left=2.8cm,right=2.8cm,top=3.5cm,bottom=3.cm]{geometry}
\usepackage[T1]{fontenc}
\usepackage{subfig}
\usepackage{graphicx}
\usepackage{tikz}
\usetikzlibrary{spy}

\usepackage{amsmath, amssymb, amsfonts, amsopn, cancel, amsthm}
\usepackage{xcolor}
\newcommand{\R}{\mathbb{R}}
\newcommand{\nfull}{n_{\Sigma}}

\usepackage{hyperref}
\newtheorem{assumption}{Assumption}
\newtheorem{theorem}{Theorem}

\newtheorem{proposition}[theorem]{Proposition}

\begin{document}

\title{TomoSelfDEQ: Self-Supervised Deep Equilibrium Learning for Sparse-Angle CT Reconstruction}
\author{Tatiana A.~Bubba\thanks{Department of Mathematics and Computer Science, University of Ferrara, Italy ({tatiana.bubba@unife.it})}\and Matteo Santacesaria\thanks{MaLGa Center, Department of Mathematics, University of Genoa, Genoa, Italy ({matteo.santacesaria@unige.it})} \and Andrea Sebastiani\thanks{Department of Physics, Computer Science and Mathematics, University of Modena and Reggio Emilia, Italy ({andrea.sebastiani@unimore.it})}}

\date{}

\maketitle

\begin{abstract}
Deep learning has emerged as a powerful tool for solving inverse problems in imaging, including computed tomography (CT). However, most approaches require paired training data with ground truth images, which can be difficult to obtain, e.g., in medical applications. We present TomoSelfDEQ, a self-supervised Deep Equilibrium (DEQ) framework for sparse-angle CT reconstruction that trains directly on undersampled measurements. We establish theoretical guarantees showing that, under suitable assumptions, our self-supervised updates match those of fully-supervised training with a loss including the (possibly non-unitary) forward operator like the CT forward map. Numerical experiments on sparse-angle CT data confirm this finding, also demonstrating that TomoSelfDEQ outperforms existing self-super\-vised methods, achieving state-of-the-art results with as few as 16 projection angles.
\end{abstract}


\section{Introduction}
In recent years, learning-based strategies for imaging inverse problems have become a popular alternative to more classical strategies, such as regularization theory or Bayesian approaches~\cite{arridge2019solving}. 
Initial deep learning methods, generally agnostic to the physical model describing an inverse problem, included convolutional neural networks trained either to directly map acquired measurements to the desired images or to post-process an initial ``standard'' reconstruction to remove noise and other artifacts~\cite{jin2017deep,lucas2018using}. More recent approaches strive to retain the information coming from the physics of the imaging methodology by incorporating it in the learned scheme. This is the case of deep unrolling or unfolding where a fixed number of iterations of an iterative scheme is regarded as layers of a deep architecture and trained end-to-end~\cite{adler2017solving,gregor2010learning}.
A particularly promising direction is the Deep Equilibrium (DEQ) paradigm \cite{gilton2021deep}, which allows unrolling an effectively infinite number of iterations by regarding minimizers of certain functionals as fixed points of suitable operators. Compared to unrolling, DEQ trains a deep implicit model, thus significantly reducing the associated \mbox{memory cost.}

While these approaches have shown impressive results, as supervised methods they typically require paired training data with ground truth images, which can be very limiting in, e.g., medical imaging applications. Self-supervised learning addresses this limitation while retaining the power of learning-based strategies. Well known methods include Plug-and-Play (PnP)~\cite{venkatakrishnan2013plug}, equivariant imaging~\cite{chen2021equivariant} and Noise2Noise~\cite{lehtinen2018noise2noise}.
Very recently, building on the latter approach, a self-supervised deep equilibrium model, called SelfDEQ, has been introduced in~\cite{gan2023self} as a framework for training model-based implicit neural networks using Jacobian-Free Backpropagation (JFB). In~\cite{gan2023self} it is also proved that, for unitary operators (e.g., magnetic resonance imaging (MRI)), SelfDEQ updates match those of fully-supervised DEQ. 

Among imaging inverse problems, computed tomography (CT) presents a particularly challenging case where learning without ground truth would be especially valuable. In sparse-angle CT, high-quality reconstruction from scarce angular measurements is crucial for reducing radiation dose in medical imaging and enabling rapid inspection in industrial applications. However, achieving such reconstructions without ground truth data remains challenging, and a framework like SelfDEQ is not directly applicable due to the non-unitary nature of the CT forward operator.

In this work, we introduce TomoSelfDEQ, a novel self-supervised deep equilibrium framework for sparse-angle CT reconstruction, which allows the following key contributions: (1) we extend the theoretical analysis of SelfDEQ to non-unitary forward operators, providing a more general relation between self- and fully-supervised losses (and corresponding JFB); (2) through extensive experiments, we demonstrate the effectiveness of TomoSelfDEQ on sparse-angle CT reconstruction, achieving state-of-the-art results even with severely undersampled data; (3) we provide theoretical guarantees establishing the equivalence between our self-supervised approach and fully-supervised training, ensuring optimal performance without the need for ground truth data.

This work fits into the line of research of training without the need for ground truth that is receiving growing interest in the imaging inverse problems community. 
In the context of tomographic imaging, several approaches have been proposed. Noise2Inverse~\cite{hendriksen2020noise2inverse}, extends the ideas of Noise2Self~\cite{batson2019noise2self} to image reconstruction in the case of fully sampled tomographic measurements. However, it does not explicitly address the challenges of sparse-angle reconstruction. More recently, Sparse2Inverse~\cite{gruber2024sparse2inverse} was proposed to specifically handle undersampled projections by combining a neural network operating in the reconstruction domain with a loss computed in the projection domain. 
In particular, Sparse2Inverse aims to address limitations of previous methods by computing the loss in the projection domain while letting the neural network operate on images. Other approaches building on similar ideas are  Proj2Proj~\cite{unal2024proj2proj}, which focuses on low-dose data by applying self-supervision principles directly in the projection domain, and Self-Supervised Denoiser Framework (SDF)~\cite{valat2024self}, which leverages pre-training on densely sampled data to enhance the image reconstruction quality from undersampled data.
TomoSelfDEQ differs from all of these approaches because it leverages the Deep Equilibrium framework, and provides theoretical guarantees for the general case of non-unitary operators, i.e., it is not limited to the CT forward operator making it well-suited to other imaging modalities that are modeled by non-unitary transformations. 

The paper is organized as follows. In Section~\ref{sec:MathsModel}, we introduce the mathematical formulation of undersampled linear inverse problems and provide a summary of relevant existing approaches.
Section~\ref{sec:DEQ} presents DEQ and introduces our method, TomoSelfDEQ.
In Section~\ref{sec:theory}, we provide our theoretical results, establishing the equivalence between self-supervised and fully-supervised training for non-unitary operators. Section~\ref{sec:NumericalExperiments} demonstrates the effectiveness of our approach through numerical experiments on sparse-angle CT reconstruction, comparing it with conventional techniques and the recently proposed Sparse2Inverse. Finally, Section~\ref{sec:conclusions} summarizes our findings and discusses future research directions.

\section{Background and mathematical formulation}\label{sec:MathsModel}

We consider linear inverse problems of the form:
\begin{equation}
    y = MAx + e,
    \label{eq:forward_model}
\end{equation}
where $y \in \R^m$ are the measurements, $e \in \R^m$ is additive white Gaussian noise (AWGN), $A \in \R^{p\times n}$ is a (non-unitary) measurement matrix, $x \in \R^n$ is the unknown target and $M \in \{0,1\}^{m\times p}$ is a matrix obtained from the identity matrix by keeping only those rows corresponding to the measured data. 
In CT, $A$ represents the Radon transform~\cite{natterer2001mathematics}, i.e., each row of $A$ corresponds to a line integral (at a specific angle) modeling the X-ray attenuation through the object of interest. As such, $A$ is a non-unitary operator. When the acquisition angles are sparse, $M$ acts as a mask selecting (from the full measurements) only those projections corresponding to the actual acquisition angles. 
Due to the ill-posedness of the Radon transform, the sparser the acquisition protocol the more challenging the reconstruction problem becomes. Therefore, traditional methods based on directly inverting the forward map, such as the filtered back-projection (FBP), generally perform poorly.
An alternative is provided by regularized approaches, which seek
$x$ by iteratively solving optimization problems of the form:
\begin{equation}\label{eq:ReguFormulation}
\hat{x} = \arg\min_{x \in \mathbb{R}^n_+}\left\{g(x) + h(x)\right\},
\end{equation}
where $g$ is the data-fidelity term that quantifies the discrepancy between the measurements and the solution and $h$ is a regularizer that imposes prior knowledge on the unknown image. In addition, in tomographic imaging it is natural to restrict the X-ray attenuation coefficients to be non-negative, that is, $\mathbb{R}^n_+$ denotes the non-negative orthant of $\mathbb{R}^n$. A common choice for the data fidelity is the least square term:
\begin{equation}\label{eq:data_fidelity}
g(x) = \frac{1}{2} \|y - MAx\|^2_2,
\end{equation}
and for the regularizer $h(x)$, Total Variation (TV) or other sparsity-promoting penalties are often employed to preserve edge-like features~\cite[Section 2]{arridge2019solving}.

While these classical approaches generally allow for accurate reconstructions from fewer tomographic measurements than usually required by FBP, they rely on hand-crafted regularizers (which can impact the mathematical tractability of the functional in \eqref{eq:ReguFormulation}), and can be sensitive to parameter choices. 

In recent years, the theoretical understanding that comes from regularized approaches has been combined with the practical advantages of learning-based methods showing that, in many instances, the hybrid approaches are able to significantly surpass both pure model- and more data-based methods retaining a sufficient degree of theoretical understanding and guarantees. Among many, we consider here Deep Equilibrium (DEQ), a paradigm initially introduced for supervised learning and very recently adapted in~\cite{gan2023self} to work as in the self-supervised framework using Jacobian-Free Backpropagation (JFB).

\section{Deep Equilibrium Models}\label{sec:DEQ}
Deep Equilibrium (DEQ)  
enables training recursive networks with effectively infinite layers without storing intermediate variables~\cite{gilton2021deep}. The key idea is to find a fixed point $\bar{x}$ of a suitable operator $T_\theta$ rather than explicitly unrolling a finite number of iterations, that is:
\begin{equation}
    \bar{x} = T_\theta(\bar{x}, y),
    \label{eq:fixed_point}
\end{equation}
where $y$ is the measurement vector and $T_\theta$ is defined as:
\begin{equation}
T_\theta(x) = \alpha f_\theta(s) + (1-\alpha)s \quad \text{with} \quad s = x - \gamma\nabla g(x),
    \label{eq:operator}
\end{equation}
being $\alpha, \gamma > 0$ hyperparameters, $f_\theta$ a (convolutional) neural network with learnable parameters $\theta$, and $g(x)$ the data fidelity term from \eqref{eq:data_fidelity}. DEQ is generally trained end-to-end in a supervised fashion, considering the empirical risk as minimization loss. In this setting, the neural network $f_\theta$ enforces a priori properties on the resulting fixed point $\bar{x}$, functioning as a denoiser similar to PnP~\cite{venkatakrishnan2013plug}.

In~\cite{gan2023self}, the authors propose to combine DEQ with the self-supervised framework of Noise2Noise, which uses different noisy realizations of the same ground truth image to train a network without the need for the actual ground truth. The resulting method, named SelfDEQ,  considers pairs of measurements $\{y_i, y'_i\}_{i=1}^N$ of the same underlying image $x_i$:
\begin{equation}
    y_i = M_i Ax_i + e_i \quad \text{and} \quad y'_i = M'_i Ax_i + e'_i
    \label{eq:measurement_pairs}
\end{equation}
where $N \geq 1$ denotes the number of training pairs  and, in~\cite{gan2023self}, $A$ is the operator associated with MRI (i.e., the Fourier transform). 
Then, training the method consists of two steps in each training iteration.
\paragraph{Forward Pass.} Starting from an initial guess $x^0$, the fixed point $\bar{x}$ is computed by running a fixed-point iteration of the form: 
\begin{equation}
    x^{k} = T_\theta(x^{k-1}, y)   
    \label{eq:forward_iteration}
\end{equation}
until convergence or a maximum number of iterations is reached. Generally, an acceleration algorithm (e.g., Anderson acceleration) is used. 
\paragraph{Backward Pass.} Given a loss function $\mathcal{L}$, parameter updates are computed via Jacobian-Free Backpropagation (JFB), which has been theoretically shown to provide valid descent directions for training implicit networks \cite{fung2022jfb}, and avoids explicitly computing and storing the inverse Jacobian, leading to efficient training with constant memory complexity: 
\begin{equation}
    \text{JFB}\mathcal{L}(\theta) =  \operatorname{Real} \bigg( (\nabla_\theta T_\theta(\bar{x}))^T\left[\frac{\partial\mathcal{L}}{\partial\bar{x}}\right]^T \bigg).
    \label{eq:jfb_update}
\end{equation}
To enable self-supervised training, the authors of~\cite{gan2023self} consider the following weighted loss:
\begin{equation}
    \mathcal{L}_\text{self}(\theta) = \mathbb{E} \left[\frac 1 2 \|M'A\bar{x}(\theta,y) - y'\|^2_W\right]
    \label{eq:self_loss}
\end{equation}
where $\|\cdot\|_W$ denotes the weighted norm given by $\|z\|^2_W = z^T W z$ for $z \in \R^m$. In~\cite{gan2023self}, $W = M' \overline W(M' \overline W)^T \in \R^{m\times m}$ is a subsampling of the diagonal weighting matrix $\overline W$, whose entries are defined by:
\begin{equation}
    \overline w_k = \begin{cases}
        \frac{1}{\sqrt{\mathbb E[M'^T M']_{k,k}}} & \text{if } \sqrt{\mathbb E[M'^T M']_{k,k}} \neq 0 \\
        0 & \text{otherwise}
    \end{cases}.
    \label{eq:weights}
\end{equation}
Note that, by minimizing $\mathcal{L}_\text{self}$, we implicitly learn the regularization through $f_{\theta}$.
The main theoretical result in~\cite{gan2023self} 
establishes that the JFB updates from the weighted self-supervised loss in~\eqref{eq:jfb_update} match those obtained using conventional supervised learning on DEQ, i.e., considering the supervised loss: 
\begin{equation} \label{eq:usual_supervised_loss}
\mathcal{L}'_\text{sup} = \mathbb{E}\left[\| \bar x - x\|_2^2   \right].
\end{equation}
However, this result was only derived under the limiting assumption of $A$ being a unitary operator. 

\subsection{TomoSelfDEQ}
The framework of SelfDEQ can be easily extended to work with more general non-unitary forward operators $A$, as it is the case of tomographic imaging. 
To this end, we introduce TomoSelfDEQ, which is particularly well-suited for sparse-angle CT reconstruction, where both the physics of the imaging process and the self-supervised nature of the training need to be carefully considered.
Starting from the original SelfDEQ framework, we modify~\eqref{eq:operator} to include the projection onto the positive orthant $\Pi_+$, that is:
\begin{equation}
    T_\theta(x) = \Pi_+(\alpha f_\theta(s) + (1-\alpha)s) \quad \text{with} \quad s = x - \gamma\nabla g(x).
    \label{eq:operatorNNeg}
\end{equation}
The measurement pairs in~\eqref{eq:measurement_pairs} are obtained by splitting each measurement acquisition into two complementary subsets of angles. Each training iteration consists of a forward and a backward pass, as for SelfDEQ. The definition of the self-supervised loss $\mathcal{L}_\text{self}(\theta)$ and the diagonal weighting matrix $\overline{W}$, which is crucial for handling the possibly non-uniform sampling density across angles, are as in~\eqref{eq:self_loss} and \eqref{eq:weights}, respectively.

Our main theoretical result is presented in the next section. In particular, Theorem~\ref{thm:losses_equivalence} establishes  that the JFB updates from the weighted self-supervised loss~\eqref{eq:self_loss} match those obtained using the following supervised loss:
\begin{equation}
    \mathcal{L}_\text{sup} = \mathbb{E}\left[\frac{1}{2}\|A(\bar{x} - x)\|_2^2\right].
    \label{eq:supervised_loss}
\end{equation}
Note that this loss differs from that of~\cite{gan2023self} (and more in general from standard supervised losses in deep learning), as it includes the forward operator $A$ in its formulation. This choice is crucial for establishing the equivalence with the self-supervised loss and reflects the physics of the imaging process. The more common supervised loss~\eqref{eq:usual_supervised_loss}
would not lead to the same theoretical guarantees due to the non-unitary nature of the Radon transform.

\section{Theoretical Results}\label{sec:theory}

We now establish the relationship between self-supervised and supervised training of TomoSelfDEQ for non-unitary operators. Our analysis relies on two key assumptions.

\begin{assumption}
The training samples correspond to the setting in \eqref{eq:measurement_pairs} with $x \sim \pi_x$, $M \sim \pi_M$, $M' \sim \pi_M$, $e \sim \mathcal{N}(0,\sigma^2I)$, and $e' \sim \mathcal{N}(0,\sigma^2I)$ drawn i.i.d. from their respective distributions.
\label{assump:independence}
\end{assumption}

\begin{assumption}
$A$ and $\mathbb{E}_M[M^T M]$, where the expectation is taken over $\pi_M$, have full rank.
\label{assump:fullrank}
\end{assumption}

Assumption \ref{assump:independence} is a mild statistical requirement stating that sampling matrices, images, and noise are sampled independently. Assumption \ref{assump:fullrank} ensures that the sampling matrices provide complete coverage of the measurement domain, though each individual matrix can still represent undersampled measurements.

Our first result establishes a key property of the weighted measurements.

\begin{proposition}\label{prop:normal_operator}
When Assumption \ref{assump:fullrank} is satisfied,
\begin{equation}\notag
    \mathbb{E}\left[(M'A)^T W M'A\right] = A^T A
\end{equation}
where the expectation is with respect to $\pi_M$ and $W = M' \overline W(M' \overline W)^T$ is defined through \eqref{eq:weights}.
\label{prop:weighting}
\end{proposition}

\begin{proof}
Since Assumption \ref{assump:fullrank} implies that $\mathbb{E}[M']_{k,k} \neq 0$,
\begin{equation}
    \overline w_k = \frac{1}{\sqrt{\mathbb{E}[M'^T M']_{k,k}}}.
    \label{eq:weight_def}
\end{equation}
Given $M'^T M' \in \{0,1\}^{p\times p}$ and $\overline W \in \R^{p\times p}$ are both diagonal matrices, we have:
\begin{align}
    \mathbb{E}[M'^T W M'] &= \mathbb{E}[M'^T M' \overline W \overline W^T M'^T M']= \overline W \overline W^T \mathbb{E}[M'^T M' M'^T M'] = I.
    \label{eq:weight_identity}
\end{align}
Therefore,
\begin{equation}
    \mathbb{E}[(M'A)^T W M'A] = A^T \mathbb{E}[M'^T W M']A = A^T A,
    \label{eq:weight_final}
\end{equation}
where the last equation follows from \eqref{eq:weight_identity}. \qed
\end{proof}

Next, our main result shows that TomoSelfDEQ updates match those of supervised training. This implies that TomoSelfDEQ can achieve the same performance as supervised training even for non-unitary operators like the Radon transform, provided the sampling pattern satisfies our assumptions.

\begin{theorem}\label{thm:losses_equivalence}
Under Assumptions \ref{assump:independence} and \ref{assump:fullrank}, the JFB update of the weighted self-supervised loss equals its supervised counterpart:
\begin{equation}
    \text{JFB}\mathcal{L}_\mathrm{self}(\theta) = \text{JFB}\mathcal{L}_\mathrm{sup}(\theta)
    \label{eq:main_result}
\end{equation}
where $ \mathcal{L}_\mathrm{sup} = \mathbb{E}\left[\frac{1}{2}\|A(\bar{x} - x)\|_2^2\right]$ and $\mathcal{L}_\mathrm{self}$ is defined in \eqref{eq:self_loss}.
\label{thm:main}
\end{theorem} 

\begin{proof}
The supervised update $\text{JFB} \mathcal{L}_{\text{sup}} (\theta)$ is given by:
\begin{equation} \notag
\text{JFB} \mathcal{L}_{\text{sup}} (\theta) = \mathbb{E} \left[ (\nabla_\theta T_\theta (\bar{x}))^T (A^T A) (\bar{x} - x) \right].
\end{equation}
For the self-supervised update, let $H' := \sqrt{W} M' A$. Then:
\begin{align}\notag
\text{JFB}\mathcal{L}_{\text{self}}(\theta) &= \mathbb{E}\left[(\nabla_\theta T_\theta(\bar{x}))^T H'^T(H' \bar{x} - \sqrt{W} y')\right] \\ \notag
&= \mathbb{E}\left[(\nabla_\theta T_\theta(\bar{x}))^T \mathbb{E}\left[H'^T(H' \bar{x} - \sqrt{W} (M'Ax + e'))\Big|x, M, e\right]\right]
\end{align}
The conditional expectation can be decomposed as:
\begin{align}\notag
\mathbb{E}\left[H'^T(H' \bar{x} - \sqrt{W} (M'Ax + e'))\Big|x, M, e\right] &= \mathbb{E}[H'^T H'](\bar{x} - x) + \mathbb{E}[H'^T\sqrt{W}]\mathbb{E}[e'] \\ \notag
&= A^T A(\bar{x} - x)
\end{align}
where we used Proposition \ref{prop:normal_operator} and $\mathbb{E}[e'] = 0$. Therefore,
\begin{equation}\notag
\text{JFB}\mathcal{L}_{\text{self}}(\theta) = \mathbb{E}\left[(\nabla_\theta T_\theta(\bar{x}))^T (A^T A)(\bar{x} - x)\right] = \text{JFB} \mathcal{L}_{\text{sup}} (\theta)
\end{equation}
which establishes the desired result. \qed
\end{proof}

\subsection{Properties of Random Sampling Patterns}

For practical implementation in sparse-angle CT, we adopt uniform random sampling from $[0,\pi)$ to satisfy Assumption \ref{assump:fullrank}. The following proposition shows that this sampling strategy leads to a simple form for $\mathbb{E}[M^T M]$, which determines the weighting matrix $W$ in our self-supervised loss.

\begin{proposition}
When the sampling patterns $M$ are drawn uniformly at random from a fixed subset of size $s$, the expectation $\mathbb{E}[M^T M]$ is a scalar multiple of the identity matrix. Specifically, 
\begin{equation}\label{eq:unifsamp}
    \mathbb{E}[M^T M] = \frac{s}{\nfull}I,
\end{equation}
where $\nfull$ is the total number of possible measurements and $s$ is the number of selected measurements.
\label{prop:uniform}
\end{proposition}
\begin{proof}
For a masking matrix $M$ with entries in $\{0,1\}$, $M^T M$ is diagonal. Under uniform random sampling with size $s$, the probability of any index being selected is $s/\nfull$, giving $\mathbb{E}[(M^T M)_{k,k}] = s/\nfull$ for diagonal entries. Off-diagonal entries are always zero by construction. Therefore equation \eqref{eq:unifsamp} follows. \qed
\end{proof}

\section{Numerical experiments}
\label{sec:NumericalExperiments}
In this section, we describe the implementation\footnote{\url{https://github.com/sedaboni/TomoSelfDEQ/}} of TomoSelfDEQ and we show some numerical results, comparing them with conventional techniques, i.e., FBP and TV regularization with non-negativity constraints (cf.~Section~\ref{sec:MathsModel}) and the recently proposed self-supervised methods Sparse2Inverse~\cite{gruber2024sparse2inverse}, tailored to sparse CT problems. For a fair comparison with our method, we applied Sparse2Inverse on simulated measurements with AWGN,  which differs from its original implementation. We conducted all the experiments on a workstation equipped with an Intel i9-12900K processor and a NVIDIA RTX A5000 GPU. 

The operator $A$ is a discretization of the Radon transform, implemented with Tomosipo \cite{hendriksen2021tomosipo} (which relies on LION to specify the CT geometry\footnote{\url{https://github.com/CambridgeCIA/LION}}). We consider a discretization of the full angular range made of $\nfull=384$ equispaced angles. 
The matrix $M$ is obtained sampling a fixed number $s$ of angles from the total 384 angles and keeping only the rows corresponding to the selected angles.
The forward pass is executed until the relative norm of the differences between iterations is less than $10^{-3}$.  In case the stopping criterion is not met, the maximum number of iterations is set to 100.
Similarly to \cite{gan2023self}, we use a U-Net~\cite{ronneberger2015u} for $f_{\theta}$, considering the spectral normalization in all layers to improve the stability of the model \cite{ryu2019plug}. We adaptively set the parameter $\gamma$ in \eqref{eq:operator} depending on the number of selected angles $s$, to ensure the operator $T_\theta$ is non-expansive. Specifically, $\gamma=\frac{1}{\widehat{\gamma}_s}$, where $\widehat{\gamma}_s$ is the spectral norm of the operator $M_sA$,  with $M_s$ being the sampling matrix obtained by selecting $s$ equispaced angles.
For the optimizer, we use a Schedule-Free version of Adam~\cite{defazio2024road}. The learning rate is set to $2*10^{-4}$, the mini-batch size is 8 and we run 2000 training epochs. 

The training dataset is generated on-the-fly, considering 32 2D slices from the CBCT Walnut dataset~\cite{der2019cone}. In detail, for each training batch the undersampled measurements  are obtained randomly sampling $s$ angles from $[0,\pi)$ with a uniform distribution.
This data augmentation strategy ensures uniform coverage of the angular range during training as required by Assumption~\ref{assump:fullrank}, enabling more robust learning. It is important to point out that this strategy is feasible since we have the full range data and it is not practical in a real medical settings. It has to be intended only for validation purposes. In contrast to the random sampling used during training, the validation set is fixed and uses equispaced angles, with the sampling mask $M_s$ selecting $s$ equispaced angles.

\begin{table}[t!]
\caption{Average PSNR and SSIM on the validation set with different sparsity levels $s$. The best results are highlighted in bold.}
\label{tab1}
\centering
\begin{tabular}{|l|cc|cc|cc|}
 \multicolumn{7}{c}{\tiny{ }}\\[-0.5em]
\cline{2-7}
\multicolumn{1}{c|}{} &  \multicolumn{2}{c|}{$s=16$} &  \multicolumn{2}{c|}{$s=32$} &  \multicolumn{2}{c|}{$s=64$}\\
\cline{2-7}
\multicolumn{1}{c|}{}  & PSNR & SSIM & PSNR & SSIM & PSNR & SSIM\\
\hline
FBP & 14.27 & 0.100 & 18.42 & 0.143 & 22.87 & 0.205\\
TV & 23.83 & 0.722 & 26.14 & 0.828 & 27.39 & 0.863\\
Sparse2Inverse & 19.21 & 0.314 & 21.14 & 0.675 & 26.80 & 0.760 \\
TomoSelfDEQ & \textbf{26.25} & \textbf{0.884} & \textbf{28.26} & \textbf{0.911} & \textbf{29.41} & \textbf{0.927}\\ 
\hline
\end{tabular}
\end{table}

\begin{figure*}[t!]
\captionsetup[subfigure]{labelformat=empty}
	\centering
    \subfloat[]{
    \begin{tabular}{l}     
	\scalebox{0.9}{
	\begin{tikzpicture}
        \clip (1.48,1.6) rectangle (-1.6,-1.6);
	\begin{scope}[spy using outlines={rectangle,cyan,magnification=2,size=1.5cm}]
	\node [name=c]{	\includegraphics[height=3cm]{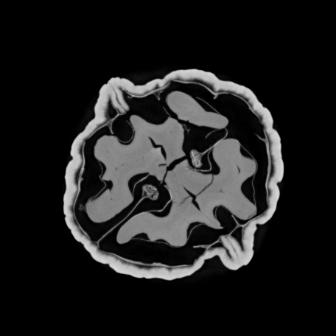}};
	\spy on (0.8,-0.2) in node [name=c1]  at (2.25,-0.75);
	\spy on (-0.5,0.2) in node [name=c1]  at (2.25,0.75);
	\end{scope}
	\end{tikzpicture}}\\
	\scalebox{0.9}{
	\begin{tikzpicture}
        \clip (1.48,1.6) rectangle (-1.6,-1.6);
	\begin{scope}[spy using outlines={rectangle,cyan,magnification=2,size=1.5cm}]
	\node [name=c]{	\includegraphics[height=3cm]{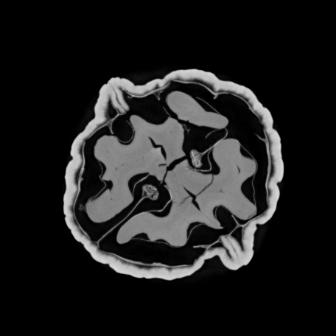}};
	\spy on (0.8,-0.2) in node [name=c1]  at (2.25,-0.75);
	\spy on (-0.5,0.2) in node [name=c1]  at (2.25,0.75);
	\end{scope}
	\end{tikzpicture}}\\
	\scalebox{0.9}{
	\begin{tikzpicture}
        \clip (1.48,1.6) rectangle (-1.6,-1.6);
	\begin{scope}[spy using outlines={rectangle,cyan,magnification=2,size=1.5cm}]
	\node [name=c]{	\includegraphics[height=3cm]{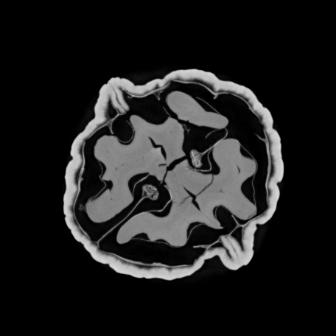}};
	\spy on (0.8,-0.2) in node [name=c1]  at (2.25,-0.75);
	\spy on (-0.5,0.2) in node [name=c1]  at (2.25,0.75);
	\end{scope}
	\end{tikzpicture}}
    \end{tabular}
    }
    \subfloat[\tiny{GT}]{
    \begin{tabular}{l}     
	\scalebox{0.9}{
	\begin{tikzpicture}
        \clip (1.48,1.6) rectangle (3.02,-1.6);
	\begin{scope}[spy using outlines={rectangle,cyan,magnification=2,size=1.5cm}]
	\node [name=c]{	\includegraphics[height=3cm]{images/experiments_sameslice/64_img_5_GT.jpg}};
	\spy on (0.8,-0.2) in node [name=c1]  at (2.25,-0.75);
	\spy on (-0.5,0.2) in node [name=c1]  at (2.25,0.75);
	\end{scope}
	\end{tikzpicture}}\\
	\scalebox{0.9}{
	\begin{tikzpicture}
        \clip (1.48,1.6) rectangle (3.02,-1.6);
	\begin{scope}[spy using outlines={rectangle,cyan,magnification=2,size=1.5cm}]
	\node [name=c]{	\includegraphics[height=3cm]{images/experiments_sameslice/32_img_5_GT.jpg}};
	\spy on (0.8,-0.2) in node [name=c1]  at (2.25,-0.75);
	\spy on (-0.5,0.2) in node [name=c1]  at (2.25,0.75);
	\end{scope}
	\end{tikzpicture}}\\
	\scalebox{0.9}{
	\begin{tikzpicture}
        \clip (1.48,1.6) rectangle (3.02,-1.6);
	\begin{scope}[spy using outlines={rectangle,cyan,magnification=2,size=1.5cm}]
	\node [name=c]{	\includegraphics[height=3cm]{images/experiments_sameslice/16_img_5_GT.jpg}};
	\spy on (0.8,-0.2) in node [name=c1]  at (2.25,-0.75);
	\spy on (-0.5,0.2) in node [name=c1]  at (2.25,0.75);
	\end{scope}
	\end{tikzpicture}}
    \end{tabular}
    }
	\subfloat[\tiny{FBP}]{
    \begin{tabular}{l}     
	\scalebox{0.9}{
	\begin{tikzpicture}
        \clip (1.48,1.6) rectangle (3.02,-1.6);
	\begin{scope}[spy using outlines={rectangle,cyan,magnification=2,size=1.5cm}]
	\node [name=c]{	\includegraphics[height=3cm]{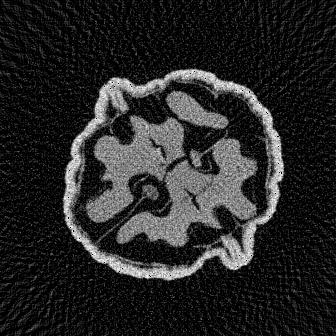}};
	\spy on (0.8,-0.2) in node [name=c1]  at (2.25,-0.75);
	\spy on (-0.5,0.2) in node [name=c1]  at (2.25,0.75);
	\end{scope}
	\end{tikzpicture}}\\
	\scalebox{0.9}{
	\begin{tikzpicture}
        \clip (1.48,1.6) rectangle (3.02,-1.6);
	\begin{scope}[spy using outlines={rectangle,cyan,magnification=2,size=1.5cm}]
	\node [name=c]{	\includegraphics[height=3cm]{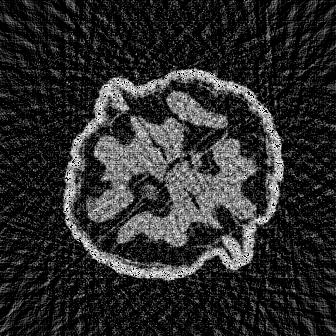}};
	\spy on (0.8,-0.2) in node [name=c1]  at (2.25,-0.75);
	\spy on (-0.5,0.2) in node [name=c1]  at (2.25,0.75);
	\end{scope}
	\end{tikzpicture}}\\
	\scalebox{0.9}{
	\begin{tikzpicture}
        \clip (1.48,1.6) rectangle (3.02,-1.6);
	\begin{scope}[spy using outlines={rectangle,cyan,magnification=2,size=1.5cm}]
	\node [name=c]{	\includegraphics[height=3cm]{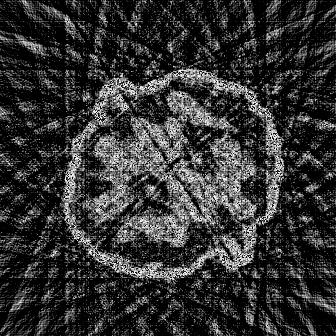}};
	\spy on (0.8,-0.2) in node [name=c1]  at (2.25,-0.75);
	\spy on (-0.5,0.2) in node [name=c1]  at (2.25,0.75);
	\end{scope}
	\end{tikzpicture}}
    \end{tabular}
    }
    \subfloat[\tiny{TV}]{
    \begin{tabular}{l}     
	\scalebox{0.9}{
	\begin{tikzpicture}
        \clip (1.48,1.6) rectangle (3.02,-1.6);
	\begin{scope}[spy using outlines={rectangle,cyan,magnification=2,size=1.5cm}]
	\node [name=c]{	\includegraphics[height=3cm]{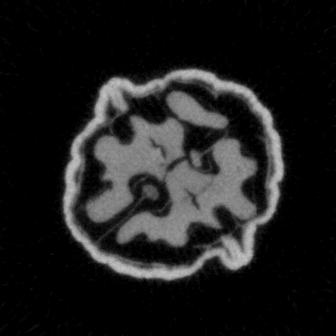}};
	\spy on (0.8,-0.2) in node [name=c1]  at (2.25,-0.75);
	\spy on (-0.5,0.2) in node [name=c1]  at (2.25,0.75);
	\end{scope}
	\end{tikzpicture}}\\
	\scalebox{0.9}{
	\begin{tikzpicture}
        \clip (1.48,1.6) rectangle (3.02,-1.6);
	\begin{scope}[spy using outlines={rectangle,cyan,magnification=2,size=1.5cm}]
	\node [name=c]{	\includegraphics[height=3cm]{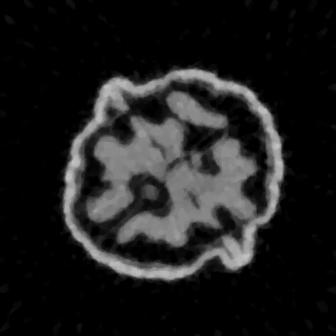}};
	\spy on (0.8,-0.2) in node [name=c1]  at (2.25,-0.75);
	\spy on (-0.5,0.2) in node [name=c1]  at (2.25,0.75);
	\end{scope}
	\end{tikzpicture}}\\
	\scalebox{0.9}{
	\begin{tikzpicture}
        \clip (1.48,1.6) rectangle (3.02,-1.6);
	\begin{scope}[spy using outlines={rectangle,cyan,magnification=2,size=1.5cm}]
	\node [name=c]{	\includegraphics[height=3cm]{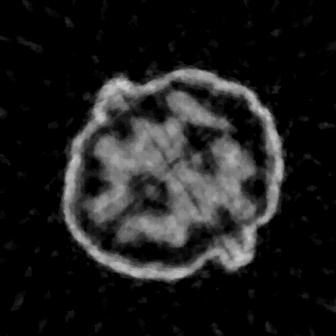}};
	\spy on (0.8,-0.2) in node [name=c1]  at (2.25,-0.75);
	\spy on (-0.5,0.2) in node [name=c1]  at (2.25,0.75);
	\end{scope}
	\end{tikzpicture}}
    \end{tabular}
    }
    \subfloat[\tiny{Sparse2Inverse}]{
    \begin{tabular}{l}     
	\scalebox{0.9}{
	\begin{tikzpicture}
        \clip (1.48,1.6) rectangle (3.02,-1.6);
	\begin{scope}[spy using outlines={rectangle,cyan,magnification=2,size=1.5cm}]
	\node [name=c]{	\includegraphics[height=3cm]{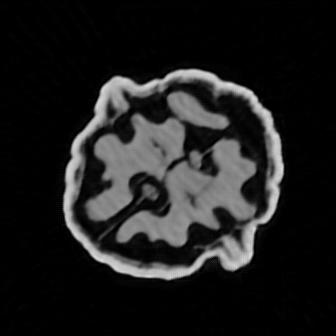}};
	\spy on (0.8,-0.2) in node [name=c1]  at (2.25,-0.75);
	\spy on (-0.5,0.2) in node [name=c1]  at (2.25,0.75);
	\end{scope}
	\end{tikzpicture}}\\
	\scalebox{0.9}{
	\begin{tikzpicture}
        \clip (1.48,1.6) rectangle (3.02,-1.6);
	\begin{scope}[spy using outlines={rectangle,cyan,magnification=2,size=1.5cm}]
	\node [name=c]{	\includegraphics[height=3cm]{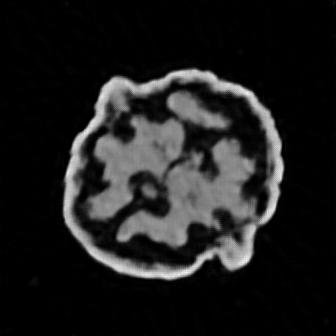}};
	\spy on (0.8,-0.2) in node [name=c1]  at (2.25,-0.75);
	\spy on (-0.5,0.2) in node [name=c1]  at (2.25,0.75);
	\end{scope}
	\end{tikzpicture}}\\
	\scalebox{0.9}{
	\begin{tikzpicture}
        \clip (1.48,1.6) rectangle (3.02,-1.6);
	\begin{scope}[spy using outlines={rectangle,cyan,magnification=2,size=1.5cm}]
	\node [name=c]{	\includegraphics[height=3cm]{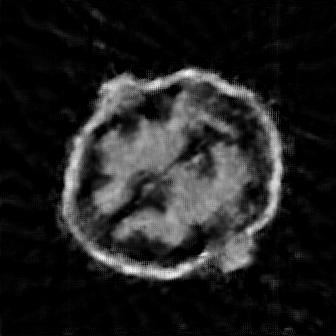}};
	\spy on (0.8,-0.2) in node [name=c1]  at (2.25,-0.75);
	\spy on (-0.5,0.2) in node [name=c1]  at (2.25,0.75);
	\end{scope}
	\end{tikzpicture}}
    \end{tabular}
    }
    \subfloat[\tiny{TomoSelfDEQ}]{
    \begin{tabular}{l}     
	\scalebox{0.9}{
	\begin{tikzpicture}
        \clip (1.48,1.6) rectangle (3.02,-1.6);
	\begin{scope}[spy using outlines={rectangle,cyan,magnification=2,size=1.5cm}]
	\node [name=c]{	\includegraphics[height=3cm]{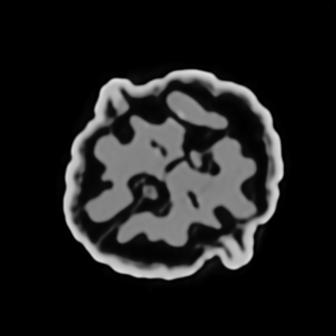}};
	\spy on (0.8,-0.2) in node [name=c1]  at (2.25,-0.75);
	\spy on (-0.5,0.2) in node [name=c1]  at (2.25,0.75);
	\end{scope}
	\end{tikzpicture}}\\
	\scalebox{0.9}{
	\begin{tikzpicture}
        \clip (1.48,1.6) rectangle (3.02,-1.6);
	\begin{scope}[spy using outlines={rectangle,cyan,magnification=2,size=1.5cm}]
	\node [name=c]{	\includegraphics[height=3cm]{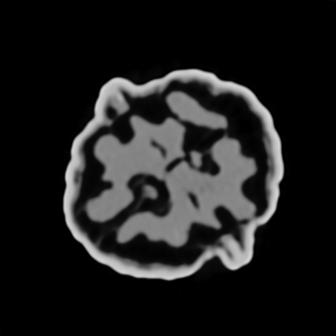}};
	\spy on (0.8,-0.2) in node [name=c1]  at (2.25,-0.75);
	\spy on (-0.5,0.2) in node [name=c1]  at (2.25,0.75);
	\end{scope}
	\end{tikzpicture}}\\
	\scalebox{0.9}{
	\begin{tikzpicture}
        \clip (1.48,1.6) rectangle (3.02,-1.6);
	\begin{scope}[spy using outlines={rectangle,cyan,magnification=2,size=1.5cm}]
	\node [name=c]{	\includegraphics[height=3cm]{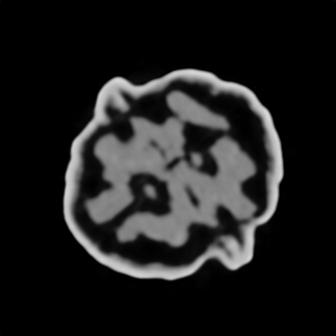}};
	\spy on (0.8,-0.2) in node [name=c1]  at (2.25,-0.75);
	\spy on (-0.5,0.2) in node [name=c1]  at (2.25,0.75);
	\end{scope}
	\end{tikzpicture}}
    \end{tabular}
    }
 \caption{Qualitative results on different sparsity levels $s$. Top row: $s=64$ angles. Middle row: $s=32$ angles. Bottom row: $s=16$ angles.}
 \label{fig:res_sameslice}
\end{figure*}

The numerical results are summarized in Table \ref{tab1}. 
These are evaluated in terms of peak signal-to-noise ratio (PSNR)  and structural similarity index (SSIM) considering three different sampling settings, with different sparsity levels $s=16$, $32, \ 64$ angles, and noise set to 1\% AWGN. It is evident that our method, TomoSelfDEQ, outperforms the competing approaches in terms of both PSNR and SSIM.
In Fig.~\ref{fig:res_sameslice}, we report the results obtained with the different methods considering 64 (top row), 32 (middle row) and 16 (bottom row) angles. We can see that with TomoSelfDEQ was able to reconstruct sharp and neat boundaries, even though some finer details are lost due to over-smoothing in certain regions. In addition, TomoSelfDEQ accurately reconstructs constant patches, without adding any significant artifacts and it proves to be stable with respect to the reduction of angles. On the other hand, FBP reconstructions are noisy and show an increasing number of streaking artifacts as the number of angles decreases. In the TV regularized reconstructions, the staircasing behavior is evident, and the sharpness of the boundaries deteriorates as the number of angles decreases. Sparse2Inverse produces reconstructions with clear artifacts, particularly when only 16 angles are considered.

\begin{figure*}[t!]
\captionsetup[subfigure]{labelformat=empty}
	\centering
    \subfloat[]{
    \begin{tabular}{l}     
	\scalebox{0.9}{
	\begin{tikzpicture}
        \clip (1.48,1.6) rectangle (-1.6,-1.6);
	\begin{scope}[spy using outlines={rectangle,cyan,magnification=2,size=1.5cm}]
	\node [name=c]{	\includegraphics[height=3cm]{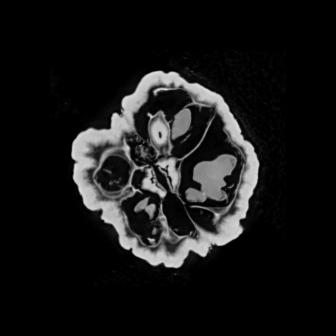}};
	\spy on (0.5,-0.2) in node [name=c1]  at (2.25,-0.75);
	\spy on (-0.5,0.2) in node [name=c2]  at (2.25,0.75);
	\end{scope}
	\end{tikzpicture}}\\
	\scalebox{0.9}{
	\begin{tikzpicture}
        \clip (1.48,1.6) rectangle (-1.6,-1.6);
	\begin{scope}[spy using outlines={rectangle,cyan,magnification=2,size=1.5cm}]
	\node [name=c]{	\includegraphics[height=3cm]{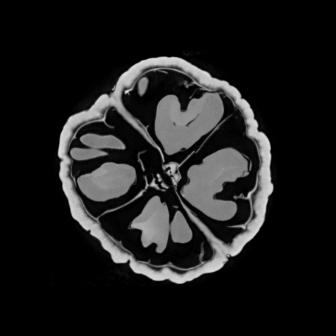}};
	\spy on (0.3,-0.4) in node [name=c3]  at (2.25,-0.75);
	\spy on (-0.1,0.4) in node [name=c4]  at (2.25,0.75);
	\end{scope}
	\end{tikzpicture}}
    \end{tabular}
    }
    \subfloat[\tiny{GT}]{
    \begin{tabular}{l}     
	\scalebox{0.9}{
	\begin{tikzpicture}
        \clip (1.48,1.6) rectangle (3.02,-1.6);
	\begin{scope}[spy using outlines={rectangle,cyan,magnification=2,size=1.5cm}]
	\node [name=c]{	\includegraphics[height=3cm]{images/experiments_sameangle/16_img_1_GT.jpg}};
	\spy on (0.5,-0.2) in node [name=c1]  at (2.25,-0.75);
	\spy on (-0.5,0.2) in node [name=c2]  at (2.25,0.75);
	\end{scope}
	\end{tikzpicture}}\\
	\scalebox{0.9}{
	\begin{tikzpicture}
        \clip (1.48,1.6) rectangle (3.02,-1.6);
	\begin{scope}[spy using outlines={rectangle,cyan,magnification=2,size=1.5cm}]
	\node [name=c]{	\includegraphics[height=3cm]{images/experiments_sameangle/16_img_26_GT.jpg}};
	\spy on (0.3,-0.4) in node [name=c3]  at (2.25,-0.75);
	\spy on (-0.1,0.4) in node [name=c4]  at (2.25,0.75);
	\end{scope}
	\end{tikzpicture}}
    \end{tabular}
    }
	\subfloat[\tiny{FBP}]{
    \begin{tabular}{l}     
	\scalebox{0.9}{
	\begin{tikzpicture}
        \clip (1.48,1.6) rectangle (3.02,-1.6);
	\begin{scope}[spy using outlines={rectangle,cyan,magnification=2,size=1.5cm}]
	\node [name=c]{	\includegraphics[height=3cm]{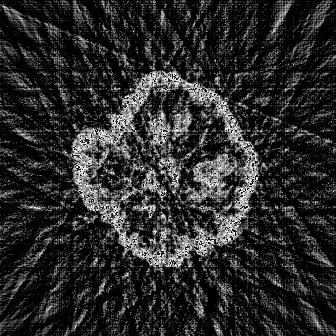}};
	\spy on (0.5,-0.2) in node [name=c1]  at (2.25,-0.75);
	\spy on (-0.5,0.2) in node [name=c2]  at (2.25,0.75);
	\end{scope}
	\end{tikzpicture}}\\
	\scalebox{0.9}{
	\begin{tikzpicture}
        \clip (1.48,1.6) rectangle (3.02,-1.6);
	\begin{scope}[spy using outlines={rectangle,cyan,magnification=2,size=1.5cm}]
	\node [name=c]{	\includegraphics[height=3cm]{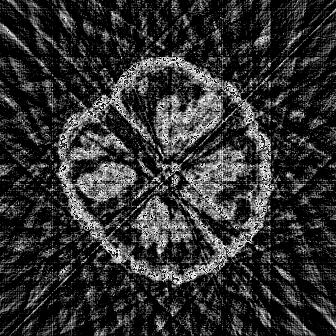}};
	\spy on (0.3,-0.4) in node [name=c3]  at (2.25,-0.75);
	\spy on (-0.1,0.4) in node [name=c4]  at (2.25,0.75);
	\end{scope}
	\end{tikzpicture}}
    \end{tabular}
    }
    \subfloat[\tiny{TV}]{
    \begin{tabular}{l}     
	\scalebox{0.9}{
	\begin{tikzpicture}
        \clip (1.48,1.6) rectangle (3.02,-1.6);
	\begin{scope}[spy using outlines={rectangle,cyan,magnification=2,size=1.5cm}]
	\node [name=c]{	\includegraphics[height=3cm]{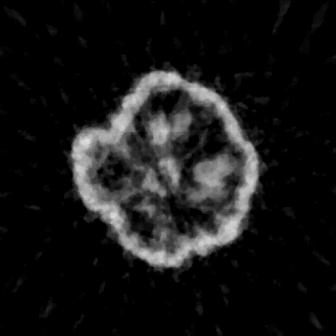}};
	\spy on (0.5,-0.2) in node [name=c1]  at (2.25,-0.75);
	\spy on (-0.5,0.2) in node [name=c2]  at (2.25,0.75);
	\end{scope}
	\end{tikzpicture}}\\
	\scalebox{0.9}{
	\begin{tikzpicture}
        \clip (1.48,1.6) rectangle (3.02,-1.6);
	\begin{scope}[spy using outlines={rectangle,cyan,magnification=2,size=1.5cm}]
	\node [name=c]{	\includegraphics[height=3cm]{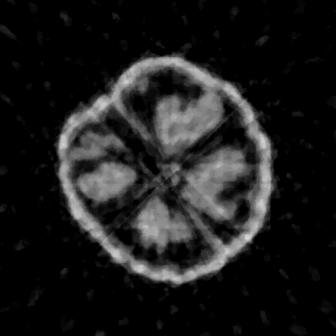}};
	\spy on (0.3,-0.4) in node [name=c3]  at (2.25,-0.75);
	\spy on (-0.1,0.4) in node [name=c4]  at (2.25,0.75);
	\end{scope}
	\end{tikzpicture}}
    \end{tabular}
    }
    \subfloat[\tiny{Sparse2Inverse}]{
    \begin{tabular}{l}     
	\scalebox{0.9}{
	\begin{tikzpicture}
        \clip (1.48,1.6) rectangle (3.02,-1.6);
	\begin{scope}[spy using outlines={rectangle,cyan,magnification=2,size=1.5cm}]
	\node [name=c]{	\includegraphics[height=3cm]{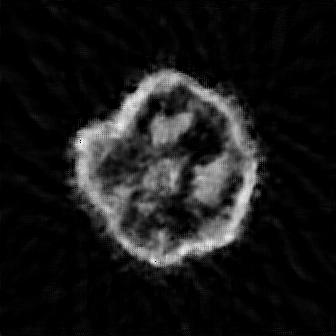}};
	\spy on (0.5,-0.2) in node [name=c1]  at (2.25,-0.75);
	\spy on (-0.5,0.2) in node [name=c2]  at (2.25,0.75);
	\end{scope}
	\end{tikzpicture}}\\
	\scalebox{0.9}{
	\begin{tikzpicture}
        \clip (1.48,1.6) rectangle (3.02,-1.6);
	\begin{scope}[spy using outlines={rectangle,cyan,magnification=2,size=1.5cm}]
	\node [name=c]{	\includegraphics[height=3cm]{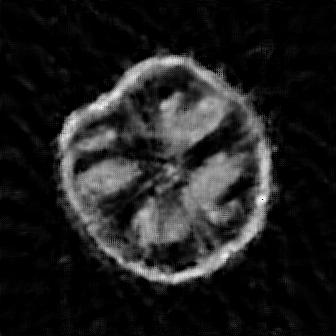}};
	\spy on (0.3,-0.4) in node [name=c3]  at (2.25,-0.75);
	\spy on (-0.1,0.4) in node [name=c4]  at (2.25,0.75);
	\end{scope}
	\end{tikzpicture}}
    \end{tabular}
    }
    \subfloat[\tiny{TomoSelfDEQ}]{
    \begin{tabular}{l}     
	\scalebox{0.9}{
	\begin{tikzpicture}
        \clip (1.48,1.6) rectangle (3.02,-1.6);
	\begin{scope}[spy using outlines={rectangle,cyan,magnification=2,size=1.5cm}]
	\node [name=c]{	\includegraphics[height=3cm]{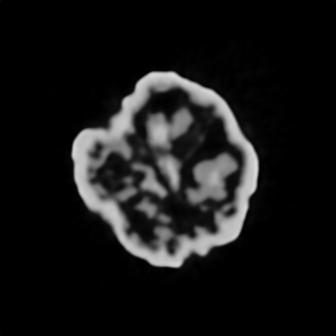}};
	\spy on (0.5,-0.2) in node [name=c1]  at (2.25,-0.75);
	\spy on (-0.5,0.2) in node [name=c2]  at (2.25,0.75);
	\end{scope}
	\end{tikzpicture}}\\
	\scalebox{0.9}{
	\begin{tikzpicture}
        \clip (1.48,1.6) rectangle (3.02,-1.6);
	\begin{scope}[spy using outlines={rectangle,cyan,magnification=2,size=1.5cm}]
	\node [name=c]{	\includegraphics[height=3cm]{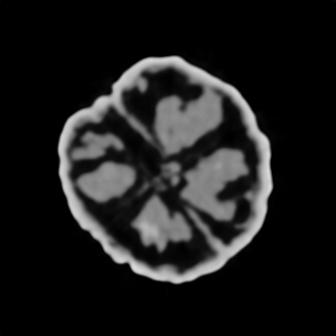}};
	\spy on (0.3,-0.4) in node [name=c3]  at (2.25,-0.75);
	\spy on (-0.1,0.4) in node [name=c4]  at (2.25,0.75);
	\end{scope}
	\end{tikzpicture}}
    \end{tabular}
    }
 \caption{Qualitative results on different walnut slices considering $s=16$ angles.}
 \label{fig:res_sameangle}
\end{figure*}

Similar consideration can be drawn for the reconstructions in Fig. \ref{fig:res_sameangle}, where we show reconstructions of different walnut slices in the most challenging setting, i.e., with 16 angles. From the close-ups it is clear that Sparse2Inverse introduces numerous noisy artifacts, compromising the reconstructions, while TomoSelfDEQ better preserves objects shapes,  significantly reducing noise presence.  A note is, however, in order on the comparison with Sparse2Inverse. While both methods share key ideas such as operating in both image and projection domains, computing losses in the projection domain, and avoiding explicit nullspace assumptions, TomoSelfDEQ differs fundamentally in several important aspects. First, while Sparse2Inverse uses a standard U-Net architecture, our method leverages DEQ to implicitly represent an infinite-depth network. Second, TomoSelfDEQ provides theoretical guarantees about matching supervised performance, while Sparse2Inverse relies primarily on empirical validation. These theoretical and architectural advantages translate into practical improvements, as demonstrated in our numerical results where TomoSelfDEQ consistently outperforms Sparse2Inverse across different undersampling rates (cf.~Table \ref{tab1}).

\begin{figure}[t!]
    \centering
    \newcommand\factor{0.3}
    \subfloat[$s=16$ angles]{
    \begin{tabular}{l}
    \includegraphics[width=\factor\textwidth]{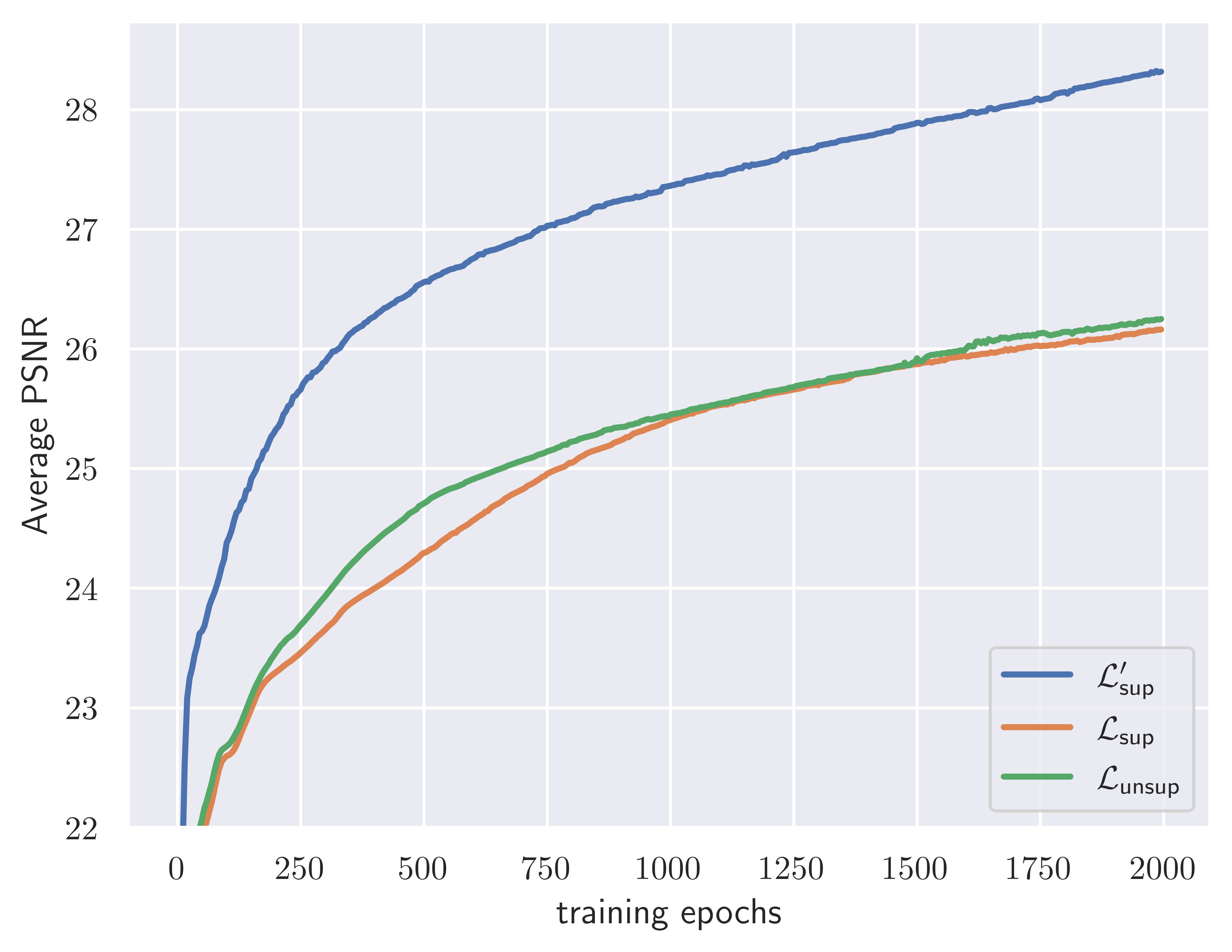}\\
    \includegraphics[width=\factor\textwidth]{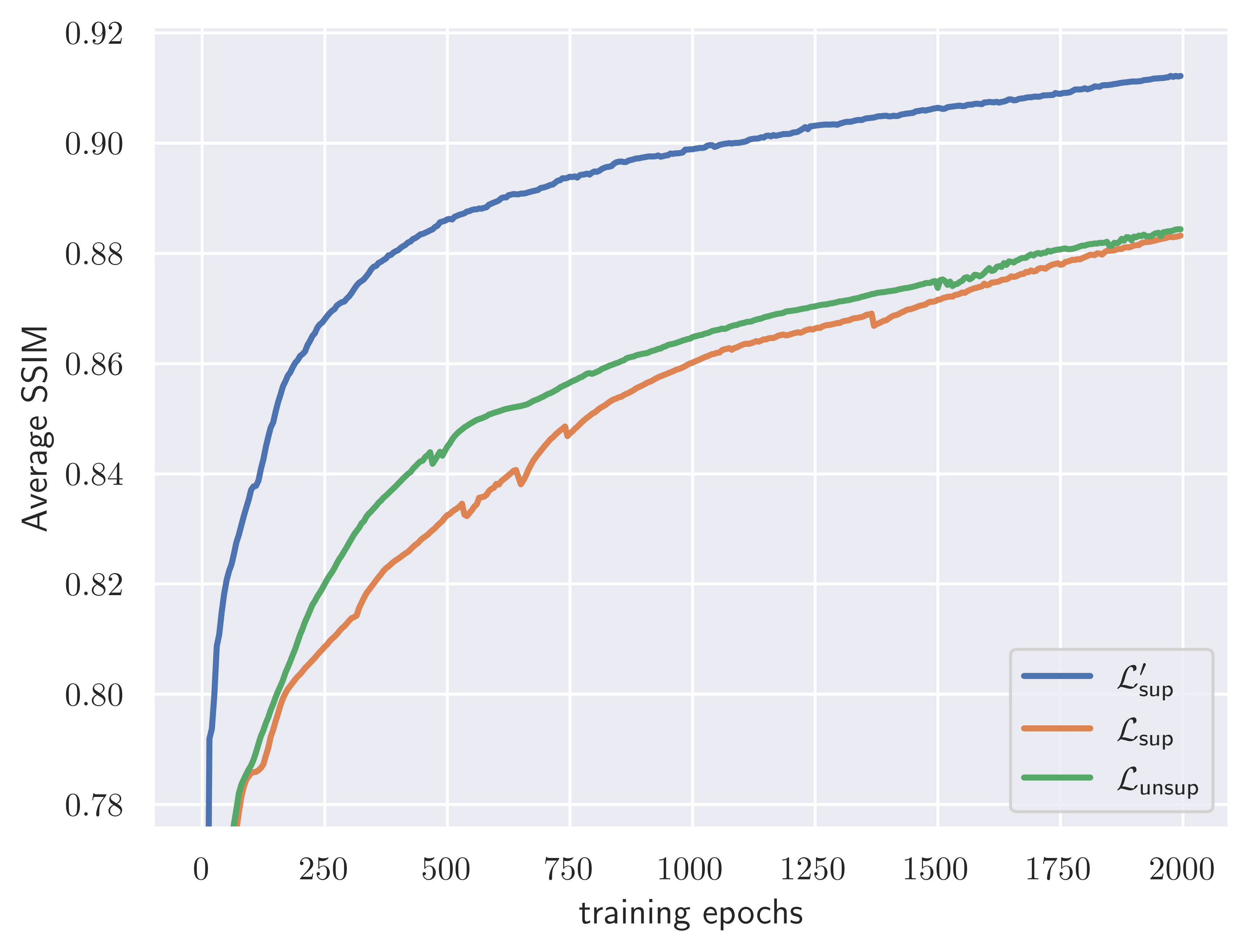}
    \end{tabular}
    }
    \subfloat[$s=32$ angles]{
    \begin{tabular}{l}
    \includegraphics[width=\factor\textwidth]{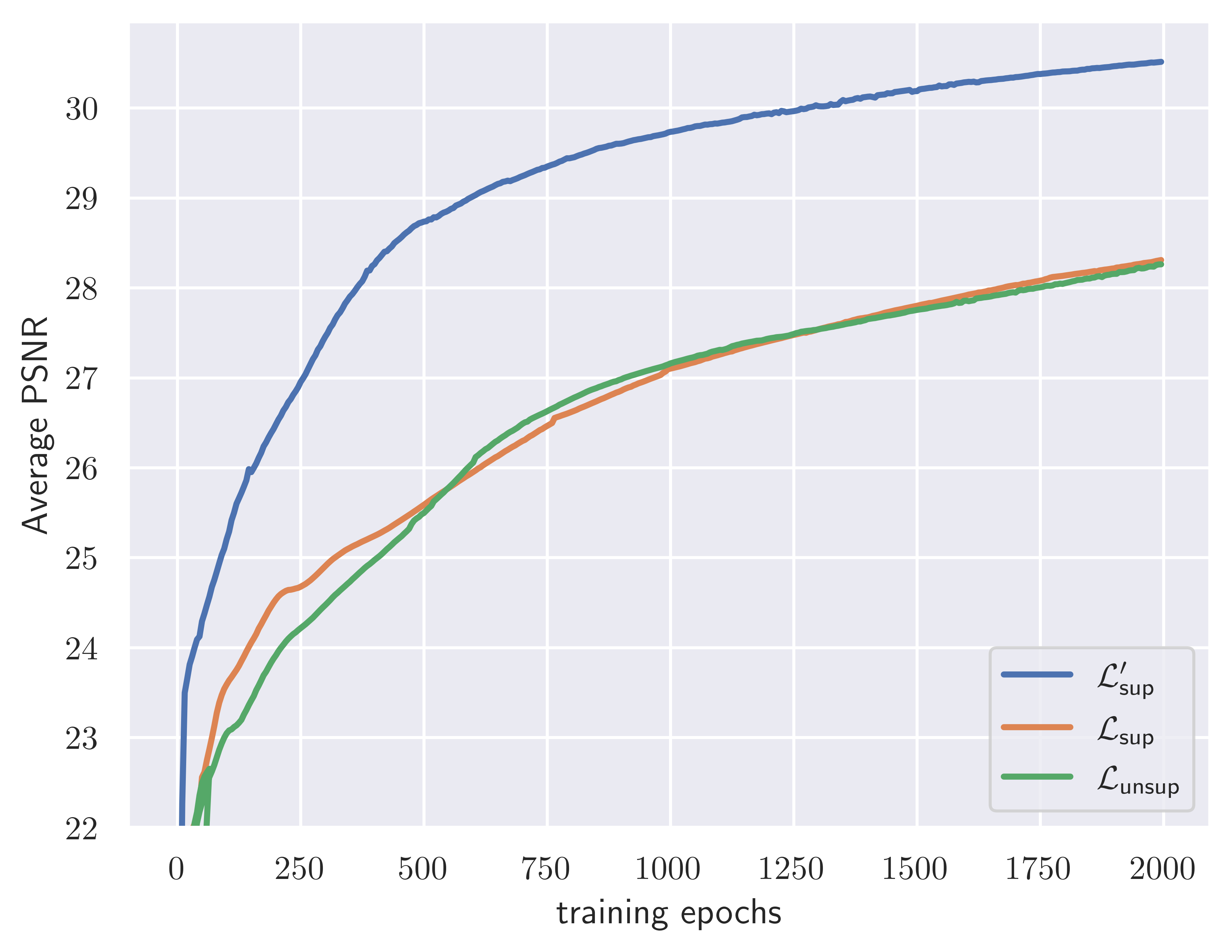}\\
    \includegraphics[width=\factor\textwidth]{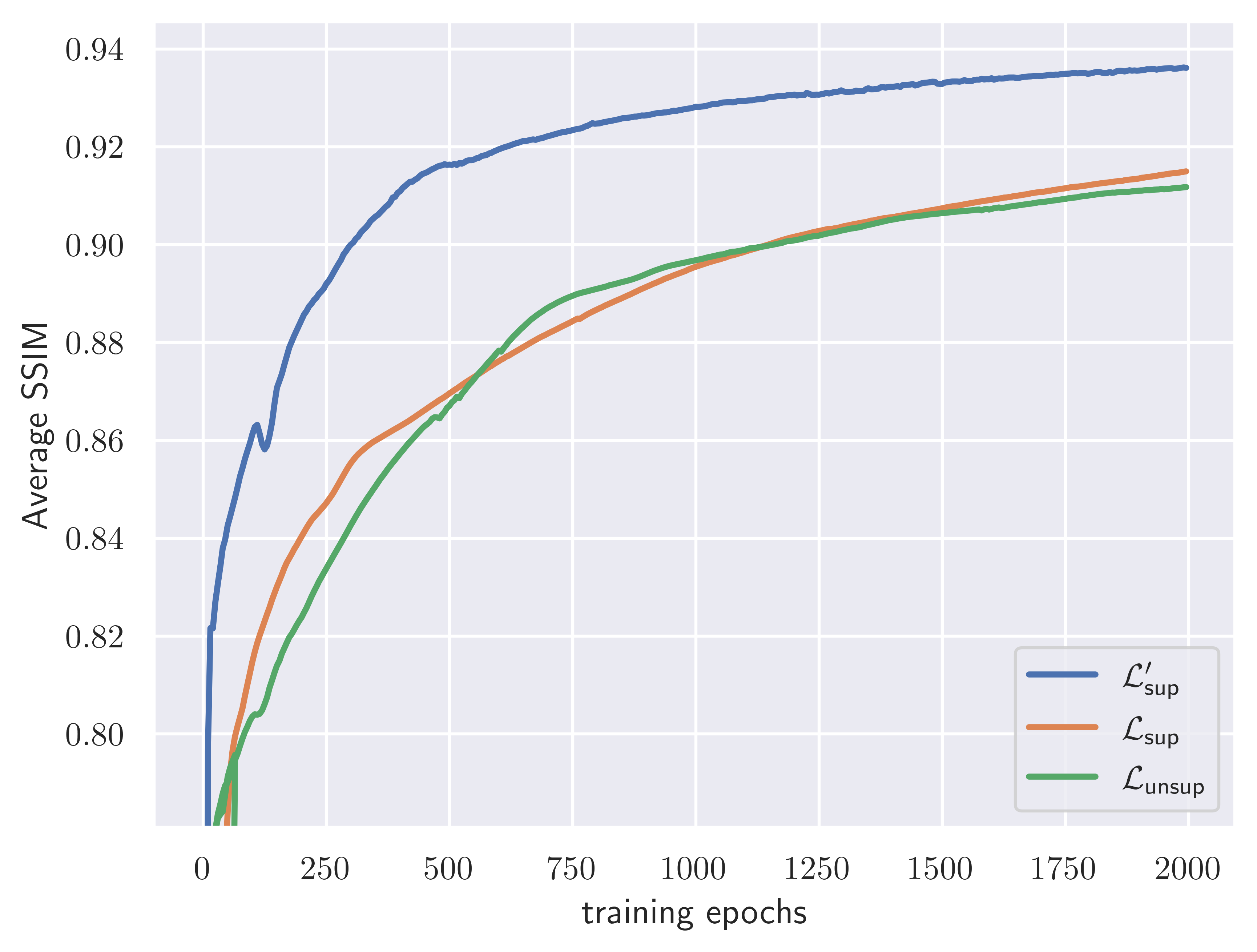}
    \end{tabular}
    }
    \subfloat[$s=64$ angles]{
    \begin{tabular}{l}
    \includegraphics[width=\factor\textwidth]{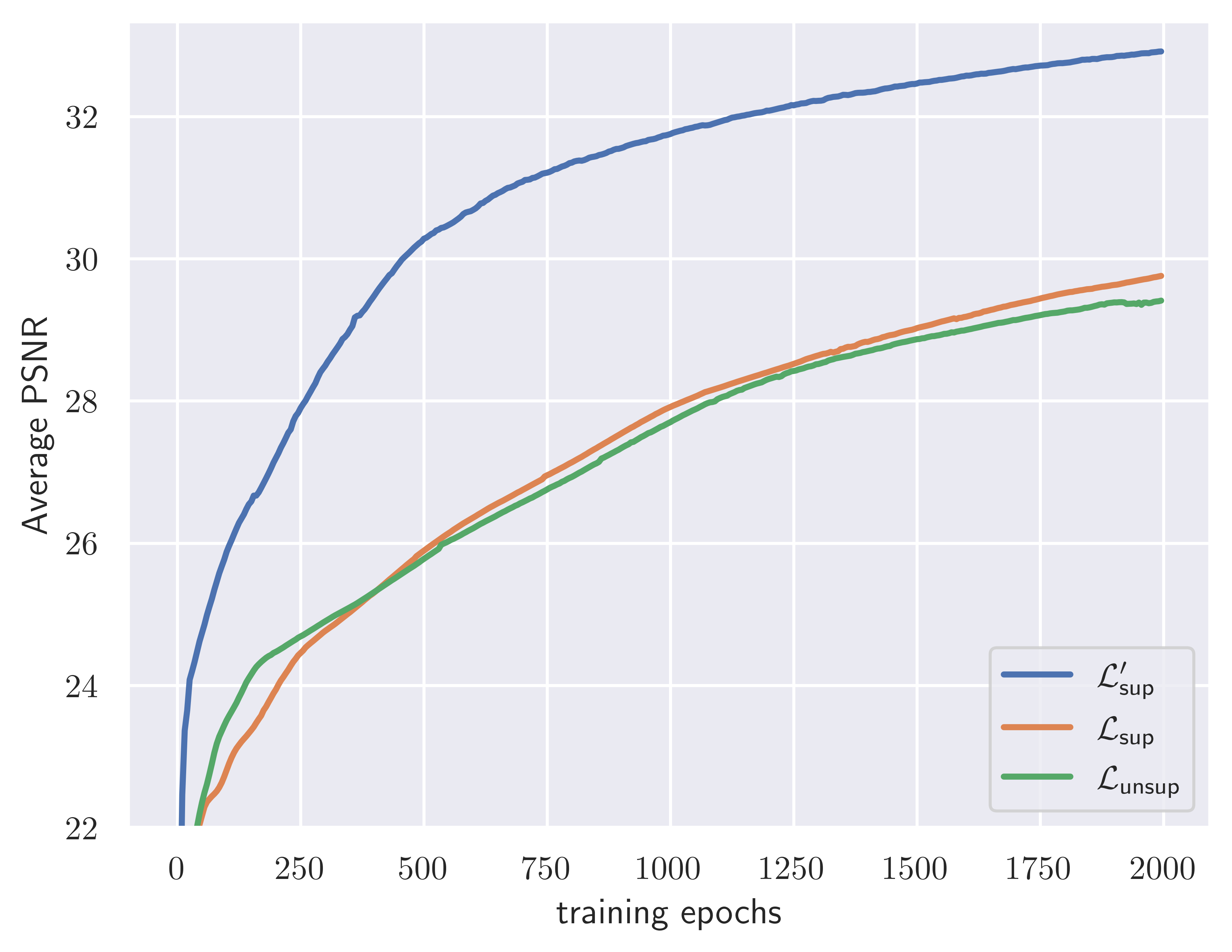}\\
    \includegraphics[width=\factor\textwidth]{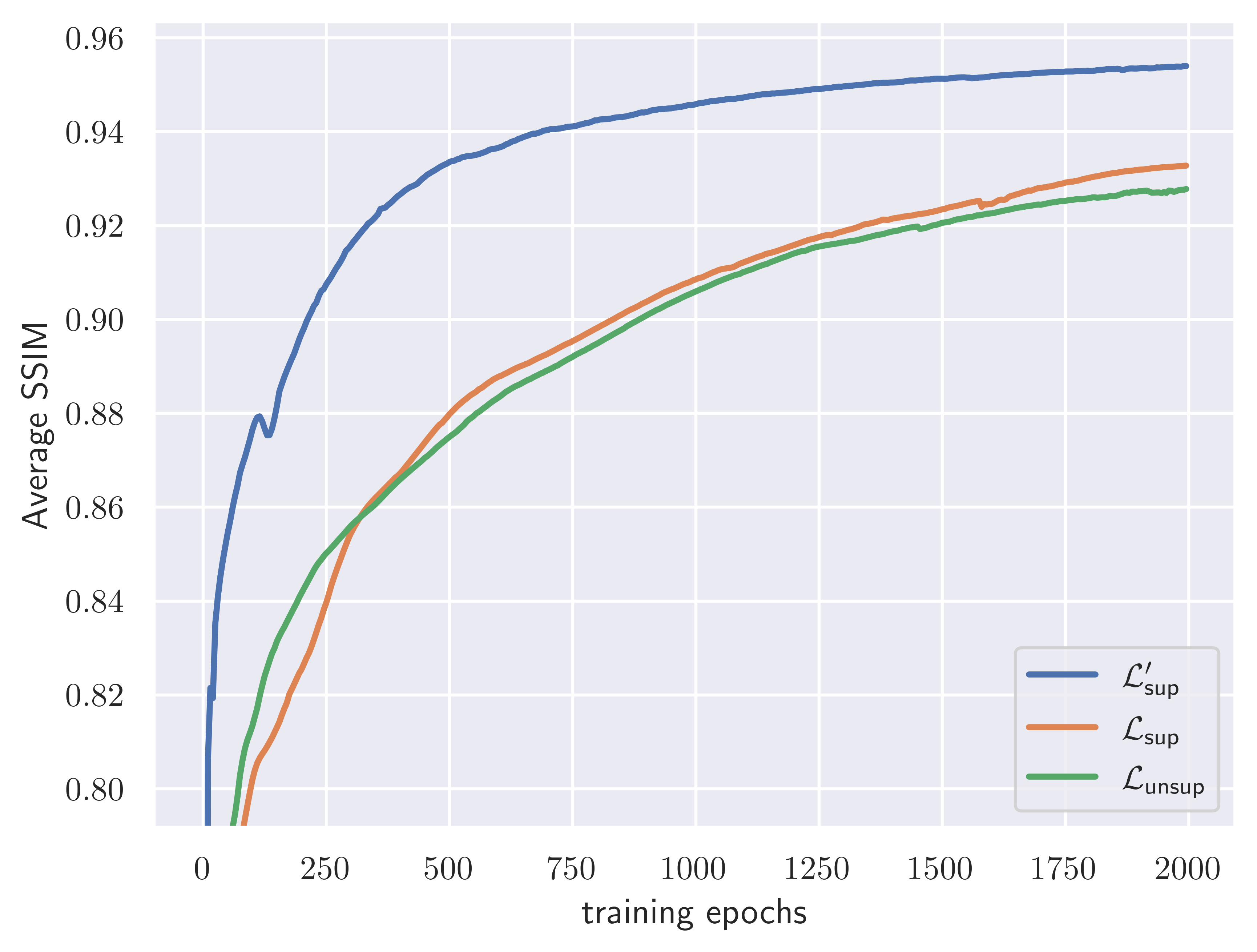}
    \end{tabular}
    }
    \caption{Quality metrics computed on the validation set and plotted with respect to the training epochs for different selection of loss. Top row: PSNR. Bottom row: SSIM.} 
    \label{fig:loss_comparison}
\end{figure}

\textbf{Ablation study.} We trained the same infinite-depth network through DEQ, considering different losses  to numerically validate the theoretical result in Theorem \ref{thm:losses_equivalence}. In particular, we compare the self-supervised loss $\mathcal{L}_\text{self}$ (cf.~\eqref{eq:self_loss}), the usual supervised loss $\mathcal{L}'_\text{sup}$ (cf.~\eqref{eq:usual_supervised_loss}), and the supervised loss $\mathcal{L}_\text{sup}$ (cf.~\eqref{eq:supervised_loss}).
We report the behavior in terms of PSNR and SSIM during training with these different losses in Fig. \ref{fig:loss_comparison}. It can be seen that the unsupervised framework achieves comparable results with that of the supervised loss $\mathcal{L}_\text{sup}$, thus confirming the result of Theorem \ref{thm:losses_equivalence}. It is evident that the results with the standard supervised loss $\mathcal{L}'_\text{sup}$ outperform the others, since in this loss the information concerning the undersampled data is not taken into account.

\section{Conclusions}
\label{sec:conclusions}
We introduced TomoSelfDEQ, a self-supervised DEQ framework for sparse-angle CT reconstruction. Our main theoretical contribution extends SelfDEQ training guarantees to non-unitary operators like the Radon transform, showing that under suitable assumptions, the self-supervised updates match those of fully-supervised training. Through extensive experiments on sparse-angle CT data, we demonstrated that TomoSelfDEQ achieves state-of-the-art reconstruction quality while requiring only undersampled measurements for training. The method consistently outperforms existing self-supervised approaches across different undersampling rates, achieving high-quality reconstructions even with as few as 16 projection angles. Future work could explore extensions to other non-unitary inverse problems and investigate the theoretical implications of different sampling strategies. Additionally, while our current validation strategy using equispaced angles demonstrates strong performance, further investigation, using different validation protocols and testing on out-of-distribution datasets, would provide valuable insights into the method's generalization capabilities.

\section*{Acknowledgments} 
The authors acknowledge (partial) support by the: Royal Society through the International Exchange scheme (grant n. IES$\backslash$R3$\backslash$223061 to TAB and MS);  European Union - NextGeneration EU (FAIR ``Future Partnership Artificial Intelligence Research'',  CUP J33C22002830006 to TAB, CUP J53C22003010006 to MS; PRIN P2022J9SNP ``Advanced optimization METhods for automated central veIn Sign detection in multiple sclerosis from magneTic resonAnce imaging" (AMETISTA) CUP E53D23017980001 to AS, PRIN 2022B32J5C ``Inverse problems in PDE: theoretical and numerical analysis", CUP D53D23005770006 to MS);  Air Force Office of Scientific Research (award number FA8655-23-1-7083 to MS); INdAM-GNCS (CUP E53C24001950001 to TAB and AS, CUP E53C23001670001 to AS); MUR Excellence Department Project (awarded to Dipartimento di Matematica, Università di Genova, CUP D33C23001110001 to MS).

\bibliographystyle{plain}
\bibliography{biblio}
\end{document}